\documentclass[showpacs,preprintnumbers,amsmath,amssymb]{revtex4}
\usepackage{amscd}
\usepackage{amsthm}
\usepackage{color}
\renewcommand{\r}{\rangle}
\def\qed{\leavevmode\unskip\penalty9999 \hbox{}\nobreak\hfill
     \quad\hbox{\leavevmode  \hbox to.77778em{%
               \hfil\vrule   \vbox to.675em%
               {\hrule width.6em\vfil\hrule}\vrule\hfil}}
     \par\vskip3pt}
\begin{document}

\title{\Large {\bf The local unitary equivalence of multipartite pure states}}

\author{Yan-Ling Wang$^{1}$, Mao-Sheng Li$^{1}$, Shao-Ming Fei$^{2,3}$, Zhu-Jun Zheng$^{1}$}

 \affiliation
 {
 {\footnotesize  {$^1$Department of Mathematics,
 South China University of Technology, Guangzhou
510640, P.R.China}} \\
{\footnotesize{
  $^2$School of Mathematical Sciences, Capital Normal University,
Beijing 100048, China}}\\
{\footnotesize{$^3$Max-Planck-Institute for Mathematics in the Sciences, 04103
Leipzig, Germany}}
}

\begin{abstract} \label{abstract}
In this paper, we give a method for the local unitary equivalent problem which is more efficient than that was proposed by Bin Liu $et \  al$ \cite{bliu}.

\bigskip
\noindent
{\bf Keywords:} local unitary equivalence, multipartite states
\end{abstract}

\pacs{03.67.Mn, 03.65.Ud}\maketitle

\maketitle

\section{Introduction}

\bigskip

Entanglement plays important roles in quantum information, such as teleportation, quantum error correction and
quantum secret sharing\cite{EofCon,nils,mhor}. In order to have a better understanding of the entanglement of quantum states, we consider their interconvertibility. That is  whether there is a local operation transfers $|\psi\r$ to $|\phi\r$ for two quantum states. Particularly, if we only consider the unitary local operation,
 the problem becomes local unitary(LU) equivalence\cite{mqubit}.

Constructing invariants of local unitary transformations is an important method to solve this problem. In \cite{Mak}, a complete set of 18 polynomial invariants is presented for qubit mixed states. In 2010, Kraus solved the LU equivalence of multipartite pure qubit states\cite{mqubit,KrausA}. Until now, the LU equivalence problem of arbitrary dimensional states have not been completely solved. Even for the arbitrary dimensional bipartite quantum mixed states, there is no good method. Recently, there are some people investigated this problem extensively\cite{bliu, zhou, zhang, Junli}. Specially in\cite{bliu}, the authors used the HOSVD decomposition  technique to reduce the LU equivalent problem to an easier one.

Bin Liu gave a method for the LU equivalence of multipartite pure states. But if the local symmetries of the HOSVD state are blocks with large size, the reduce problem is still difficult to tackle with. Our paper give a refinement of Bun Liu's method for such states. One can find that our method is more efficient than Bin Liu's through the two given examples.

In this paper, we first give a necessary condition for the LU equivalence of two quantum states. After considering the simultaneously unitary equivalence, we give a reduce form with local symmetry in smaller size than in\cite{bliu,Junli}.

This paper is organized as follows. In section II, we give some knew conclusions of the unitary equivalence and some notations of HOSVD decomposition. In section III, we illustrate our method for the tripartite pure states and give some examples. In section IV, we generalize our method to multipartite pure states and give a reduce form whose symmetry group can be made into smaller blocks than the the method\cite{bliu}.

\bigskip
\section{ Some preparations }

\bigskip
\noindent{\bf Definition \ 1.}\cite{Heydar} Let $n\in\mathbb{N}$, $r_1, r_2, ..., r_m\in \mathbb{N}^{+}$, such that
$n=\sum\limits_{i=1}^{m}r_i$ and $\{E_i\}_{i=1}^s $ be a partition of $\{1, 2,..., m\}.$
The set $ \{H=diag(U_1, U_2, ..., U_m)| U_i \in U(r_i), U_i=U_j$, if $\exists  1\leq r\leq s, i, j \in E_r\}$
is called a direct group of $M_n(\mathbb{C})$ with size $\{r_i\}_{i=1}^m$ and restricted by $\{E_i\}_{i=1}^s$.

\bigskip
\noindent{\bf  Lemma \ 1.}\cite{Little, Heydar, Benner} Let $H$ be a direct group of $n\times n $ matrix. Then $A, B$ are unitary equivalence under $ H$ if and only if they give the same reduce form $ \widetilde{A}, \widetilde{B}.$
 Moreover, the invariant group $ \{U\in H|U \widetilde{A} U^\dag=\widetilde{A}\} $ is a direct subgroup of $H$.

\bigskip
\noindent{\bf  Corollary \ 1.} Given two order matrices sets
$\{A_i\}_{i=1}^m$,  $\{B_i\}_{i=1}^m$ $\in$ $M_n(\mathbb{C})$.
The following two statements are equivalent. \\
(I) $\exists U\in U(n), $ s.t. $B_i=U A_i U^\dag, i=1, 2, ..., m;$\\
(II)$ A=diag(A_1, ..., A_m), B=diag(B_1, ..., B_m)$\\
 have the same reduce form under the direct group defined by $\{r_i=n\}_{i=1}^m, E_1=\{1, 2, ..., m\}$.

\bigskip
Now let us brief the high order singular value decomposition(HOSVD) technique which can be seen\cite{Junli, bliu}.

\bigskip
\noindent{\bf Definition \ 2.} A multipartite pure state $|\psi\rangle $ in
$$\mathbb{C}^{I_1}\otimes \mathbb{C}^{I_2}\otimes...\otimes \mathbb{C}^{I_N}$$
is called to be a HOSVD state if $ |\psi\rangle_m|\psi\rangle_m^\dag  $ are diagonal for
$m=1, ..., N$.

Suppose
$\lambda_1^{(m)}>\lambda_2^{(m)}>\ldots>\lambda_{t^{(m)}}^{(m)}\geq0$ are distinct m-mode singular values of $|\psi\rangle$ with respective positive multiplicities $\mu_1^{(m)}, \mu_2^{(m)}, \ldots, \mu_{t^{(m)}}^{(m)}$ where $\sum\limits_{k=1}^{t^{(m)}}\mu_k^{(m)}=I_m$.
 In this case, if $\Omega$ is a HOSVD state of $|\psi\rangle$, then
$$\bigotimes_{m=1}^{N}(\bigoplus_{k=1}^{t^{(m)}}{U_k^{(m)}})\Omega\equiv\bigotimes_{m=1}^{N}S^{(m)}\Omega$$
is also HOSVD state of $|\psi\rangle$. Here $U_k^{(m)}\in M_{\mu_k^{(m)}}(\mathbb{C})$
is an arbitrary $\mu_k^{(m)}\times \mu_k^{(m)}$ unitary matrix and constitute the diagonal blocks
of $S^{(m)}$ which are conformal to those m-mode singular values of $|\psi\rangle$ with multiplicity.

Through out this paper, when we said a state $|\psi\rangle$ is its HOSVD state, we mean
$ |\psi\rangle_m|\psi\rangle_m^\dag  $ are diagonal with $\underbrace{\lambda_1^{(m)}, ..., \lambda_1^{(m)}}_{\mu_1^{(m)}}, ..., \underbrace{\lambda_{t^{(m)}}^{(m)}, ...,
\lambda_{t^{(m)}}^{(m)}}_{\mu_{t^{(m)}}^{(m)}}, m=1, 2 , ..., N$
and with local symmetry $\bigotimes_{m=1}^{N}S^{(m)}$.

\bigskip
\section{ Tripartite states}
\bigskip

In this section, we first consider the  tripartite pure states. If $|\psi\rangle$ is a tripartite pure state. We can suppose that it's HOSVD, then
$$|{\psi}\rangle=\sum\limits_{k=1}^{t^{(1)}}{\sum\limits_{l=1}^{\mu_k^{(1)}}{\lambda_{k}^{(1)}}}|{n_l^{(1), k}\rangle_1\otimes|{v_{l}^{(1), k}}\rangle}_{\neg 1}$$
$$\ \ \ \ =\sum\limits_{k=1}^{t^{(2)}}{\sum\limits_{l=1}^{\mu_k^{(2)}}{\lambda_{k}^{(2)}}}|{n_l^{(2), k}\rangle_2\otimes|{v_{l}^{(2), k}}\rangle}_{\neg 2}$$
\begin{equation}\label{Equ:9} \ \ =\sum\limits_{k=1}^{t^{(3)}}{\sum\limits_{l=1}^{\mu_k^{(3)}}{\lambda_{k}^{(3)}}}|{n_l^{(3), k}\rangle_3\otimes|{v_{l}^{(3), k}}\rangle}_{\neg 3},
\end{equation}
where $n_l^{(i),k}\doteq l+\sum\limits_{s=1}^{k-1}{\mu_s^{(i)}}$ and$|{v_{l}^{(i), k}}\rangle_{\neg i}$ is just the normal vector when we collect the $n_l^{(i), k}$-th term of the i-th system of $|\psi\r$.

So $|\psi\rangle$ can be looked as the purify of the mixed bipartite states
\begin{equation}\label{Equ:10}
\rho^{(1)}=\sum\limits_{k=1}^{t^{(1)}}{{\lambda_{k}^{(1)}}^{2}}{\sum\limits_{l=1}^{\mu_k^{(1)}}|{v_{l}^{(1), k}\rangle_{\neg1 \neg1}\langle{v_{l}^{(1),k}|}}},
\end{equation}
\begin{equation}\label{Equ:11}
\rho^{(2)}=\sum\limits_{k=1}^{t^{(2)}}{{\lambda_{k}^{(2)}}^{2}}{\sum\limits_{l=1}^{\mu_k^{(2)}}|{v_{l}^{(2), k}\rangle_{\neg2 \neg2}\langle{v_{l}^{(2), k}|}}},
\end{equation}
\begin{equation}\label{Equ:12}
\rho^{(3)}=\sum\limits_{k=1}^{t^{(3)}}{{\lambda_{k}^{(3)}}^{2}}{\sum\limits_{l=1}^{\mu_k^{(3)}}|{v_{l}^{(3), k}\rangle_{\neg3 \neg3}\langle{v_{l}^{(3), k}|}}}.
\end{equation}

Denote $$|\psi ^{(i), k}\rangle\doteq{\lambda_{k}^{(i)}}\sum\limits_{l=1}^{\mu_k^{(i)}}|{n_l^{(i), k}\rangle_i\otimes|{v_{l}^{(i), k}}
\rangle}_{\neg i}.$$
This can be viewed as a new tripartite pure state, even though it's not normal. In fact it is just a purify state of the $\neg i$ bipartite mixed state
\begin{equation}\label{Equ:13}
\rho_k^{(i)}={{\lambda_{k}^{(i)}}^{2}}{\sum\limits_{l=1}^{\mu_k^{(i)}}|{v_{l}^{(i), k}\rangle_{\neg i \neg i}\langle{v_{l}^{(i), k}|}}}.
\end{equation}
Then we can consider its m-mode matrix
$|\psi ^{(i), k}\rangle_m$, and we denote
$$ M_{\psi, m}^{i, k}\doteq|\psi ^{(i), k}\rangle_m|\psi ^{(i), k}\rangle_m^\dag,$$
$$1\leq m\neq i\leq 3, k=1, 2, ..., t^{(i)}.$$

The upper label $(i)$  means the i-th system as a new system to purify the $\neg i$ bipartite mixed state $\rho_k^{(i)}$, where $\rho_k^{(i)}$ is the k-th part  of $\rho^{(i)}$. After purifying the mixed state  $\rho_k^{(i)}$, we get a pure tripartite state $|\psi ^{(i), k}\r$. Considering the m-mode decomposition of the new state, the matrix $M_{\psi, m}^{i,k}$ is obtained.

\bigskip
\noindent{\bf Theorem \ 1.} Let $|\psi\rangle, |\phi\rangle$ be two HOSVD pure state in
$\mathbb{C}^{I_1}\otimes\mathbb{C}^{I_2}\otimes\mathbb{C}^{I_3}$.
If
\begin{equation}\label{Equ:3}
U_1\otimes U_2\otimes U_3|\psi\rangle=|\phi\rangle,
\end{equation}
then
$$ U_m M_{\psi, m}^{i, k} U_m^\dag=M_{\phi, m}^{i, k}, 1\leq m\neq i\leq 3, k=1, 2, ..., t^{(i)}.$$

\bigskip
\noindent{ Proof:} For HOSVD states
\begin{equation}\label{Equ:1}
|{\psi}\rangle=\sum\limits_{k=1}^{t^{(i)}}{\sum\limits_{l=1}^{\mu_k^{(i)}}{\lambda_{k}^{(i)}}}|{n_l^{(i), k}\rangle_i, \otimes|{v_{l}^{(i), k}}\rangle}_{\neg i},
\end{equation}
\begin{equation}\label{Equ:2}
|{\phi}\rangle=\sum\limits_{k=1}^{t^{(i)}}{\sum\limits_{l=1}^{\mu_k^{(i)}}{\lambda_{k}^{(i)}}}|{n_l^{(i), k}\rangle_i
\otimes|{w_{l}^{(i), k}}\rangle}_{\neg i},
\end{equation}

since $U_1\otimes U_2\otimes U_3|\psi\rangle=|\phi\rangle$,
$U_i \in S^{(i)}$ is block unitary matrix with size $\{\mu_k^{(i)}\}_{k=1}^{t^{(i)}}$.

Suppose
\begin{equation}\label{Equ:4}
U_i =\bigoplus_{k=1}^{t^{(i)}}{U_k^{(i)}}.
\end{equation}
From equations (\ref{Equ:1})(\ref{Equ:2})(\ref{Equ:3})(\ref{Equ:4}), we have
$$ U_k^{(i)}\otimes(U_1\otimes U_2\otimes U_3)_{\neg i}\sum\limits_{l=1}^{\mu_k^{(i)}}{\lambda_{k}^{(i)}}|n_l^{(i), k}\rangle_i\otimes|{v_{l}^{(i), k}}\rangle_{\neg i}
=\sum\limits_{l=1}^{\mu_k^{(i)}}{\lambda_{k}^{(i)}}|n_l^{(i), k}\rangle_i\otimes|{w_{l}^{(i), k}}\rangle_{\neg i}.$$

That is
\begin{equation}\label{Equ:important}
 U_k^{(i)}\otimes(U_1\otimes U_2\otimes U_3)_{\neg i}|\psi_k^{(i)}\rangle=|\phi_k^{(i)}\rangle.
 \end{equation}
Here $(U_1\otimes U_2\otimes U_3)_{\neg i} $ represents the matrix which gain by delete $U_i$ from $U_1\otimes U_2\otimes U_3$, the same meaning below for $( U_k^{(i)}\otimes(U_1\otimes U_2\otimes U_3)_{\neg i})_{\neg m}^t$.

Considering the m-mode of the new states, we have
$U^{(m)}|\psi_k^{(i)}\rangle_m( U_k^{(i)}\otimes(U_1\otimes U_2\otimes U_3)_{\neg i})_{\neg m}^t=|\phi_k^{(i)}\rangle_m$.
Hence
$U^{(m)}|\psi_k^{(i)}\rangle_m|\psi_k^{(i)}\rangle_m^\dag U^{(m)\dag}=|\phi_k^{(i)}\rangle_m|\phi_k^{(i)}\rangle_m^\dag$.
\qed
\medskip
\noindent{\bf Remark:} Equation(\ref{Equ:important}) is the core of the proof. The LU equivalence of two quantum states implies the LU transformation itself satisfies lots of simultaneously unitary equivalent conditions.
\medskip

In spire of  theorem 1, for $1\leq m\leq 3$, there are two groups with $L_m=\mathbb{}\sum\limits_{i=1, i\neq m}^3 t^{(i)}$ ordered matrices in $M_{I_m}(\mathbb{C})$ corresponding to the LU equivalent states $|\psi\r, |\phi\r$. Moreover, they  are simultaneously unitary equivalence.\\
Denote
$$M_{\psi, 1}=diag(M_{\psi, 1}^{2, 1}, ..., M_{\psi, 1}^{2, t^{(2)}}, M_{\psi, 1}^{3, 1}, ..., M_{\psi, 1}^{3, t^{(3)}}),$$
$$M_{\psi, 2}=diag(M_{\psi, 2}^{1, 1}, ..., M_{\psi, 2}^{1, t^{(1)}}, M_{\psi, 2}^{3, 1}, ..., M_{\psi, 2}^{3, t^{(3)}}),$$
$$M_{\psi, 3}=diag(M_{\psi, 3}^{1, 1}, ..., M_{\psi, 3}^{1, t^{(1)}}, M_{\psi, 3}^{2, 1}, ..., M_{\psi, 3}^{2, t^{(2)}}).$$

Let $H_m^M$ be the direct group of $M_{L_m\times I_m}(\mathbb{C})$ defined by size $ \{r_k=I_m\}_{k=1}^{L_m}$ and restricted by $E_1=\{1,2,...,L_m\}$,
and $H_m$ be the set of the $I_m$-th sequential principal minors of the matrices in $H_m^M$. Obviously, $H_m$ is a direct group of $M_{I_m}(\mathbb{C})$. Then the matrices in $H_m^M$ are just ${L_m}$ copies of the matrices in
$H_m$. That is, the matrices in $ H_m^M$ are just of the form $diag(\underbrace{{H}_m, ..., {H}_m}_{L_m})$.

Given a matrix $M_{\psi, m} $ by the algorithm in \cite{Heydar}, there is a matrix $U_{\psi, m}^{\widetilde{\psi}, M}$ transferring $M_{\psi, m} $ into its canonical form $M_{{\psi}, m}^0 $
with invariant subgroup $ \widetilde{H}_m^M=diag(\underbrace{\widetilde{H}_m, ..., \widetilde{H}_m}_{L_m})$.
Suppose
$$U_{\psi, m}^{\widetilde{\psi}, M}= diag(\underbrace{U_{\psi, m}^{\widetilde{\psi}, m}, ..., U_{\psi, m}^{\widetilde{\psi}, m}}_{L_m})\in H_m^M$$
with $U_{\psi, m}^{\widetilde{\psi}, m}$ in the $I_m$-th sequential principal minor.

Calculated all the three unitary matrices $\{U_{\psi, m}^{\widetilde{\psi}, m}; m=1, 2, 3\}$, a state $|\widetilde{\psi}\rangle$ can be defined as
$$|\widetilde{\psi}\rangle \doteq U_{\psi, 1}^{\widetilde{\psi}, 1}\otimes U_{\psi, 2}^{\widetilde{\psi}, 2}\otimes U_{\psi, 3}^{\widetilde{\psi}, 3}|\psi\rangle.$$

$|\widetilde{\psi}\rangle$ is called a reduce form of $|\psi \rangle$.

\bigskip
\noindent{\bf Theorem  \ 2.} Let $|\psi\rangle, |\phi\rangle$ be HOSVD states.
Then $|\psi\rangle$ and $|\phi\rangle$ are LU equivalent if and only if $|\widetilde{\psi}\rangle$
can be transferred to $|\widetilde{\phi}\rangle$ under $\widetilde{H}_1\otimes \widetilde{H}_2 \otimes \widetilde{H}_3$.

\bigskip
\noindent{\large Proof:}
If
\begin{equation}\label{Equ:5}
U_{\psi, 1}^{\phi, 1}\otimes U_{\psi, 2}^{\phi, 2}\otimes U_{\psi, 3}^{\phi, 3}|\psi\rangle=|\phi\rangle,
\end{equation}
by
the definitions of $|\widetilde{\psi}\rangle$ and $|\widetilde{\phi}\rangle$,
we have
{
\begin{equation}\label{Equ:8}
(U_{\phi, 1}^{\widetilde{\phi}, 1}U_{\psi, 1}^{\phi, 1}{U_{\psi, 1}^{\widetilde{\psi}, 1}}^\dag)\otimes
(U_{\phi, 2}^{\widetilde{\phi}, 2}U_{\psi, 2}^{\phi, 2}{U_{\psi, 2}^{\widetilde{\psi}, 2}}^\dag)\otimes
(U_{\phi, 3}^{\widetilde{\phi}, 3}U_{\psi, 3}^{\phi, 3}{U_{\psi, 3}^{\widetilde{\psi}, 3}}^\dag)
 |\widetilde{\psi}\rangle =|\widetilde{\phi}\rangle.
 \end{equation}
 }
From the four states $|\psi\r$,$|\phi\r$, $|\widetilde{\psi}\r$, $|\widetilde{\phi}\r$,
we can construct $M_{\psi, m}$, $M_{\phi, m}$, $M_{\widetilde{\psi}, m}$, $M_{\widetilde{\phi}, m}$ respectively.\\
Then there is a commutative diagram
\[\begin{CD}
M_{\psi, m} @>U_{\psi, m}^{\phi, M}>>M_{\phi, m}\\
@V U_{\psi, m}^{\widetilde{\psi}, M} VV @VV U_{\phi, m}^{\widetilde{\phi}, M} V \\
M_{\widetilde{\psi}, m} @>>U_{\widetilde{\psi}, m}^{\widetilde{\phi}, M} >M_{\widetilde{\phi}, m}
\end{CD}
\]\\
where $$U_{\widetilde{\psi}, m}^{\widetilde{\phi}, M}=diag(\underbrace{U_{\phi, m}^{\widetilde{\phi}, m}U_{\psi, m}^{\phi,m}{U_{\psi, m}^{\widetilde{\psi}, m}}^\dag,...,U_{\phi, m}^{\widetilde{\phi}, m}U_{\psi, m}^{\phi, m}{U_{\psi, m}^{\widetilde
{\psi}, m}}^\dag}_{L_m}).$$
The canonical form $M_{\psi,m}^0, M_{\phi, m}^0$ of  $M_{\psi, m}, M_{\phi, m}$ are just $M_{\widetilde{\psi}, m},M_{\widetilde{\phi}, m}.$
So $U_{\widetilde{\psi}, m}^{\widetilde{\phi}, M}$ transfers $M_{\psi, m}^0$ to $M_{\phi, m}^0$.\\
By lemma 1, we have
$$U_{\widetilde{\psi}, m}^{\widetilde{\phi}, M}\in \widetilde{H}_m^M.$$
That is,
$$(U_{\phi, m}^{\widetilde{\phi}, m}U_{\psi, m}^{\phi, m}{U_{\psi, m}^{\widetilde{\psi}, m}}^\dag)\in \widetilde{H}_m, m=1, 2, 3.$$
By equation (\ref{Equ:8}), $|\widetilde{\psi}\r$ can be transferred to $|\widetilde{\phi}\r$ under $\widetilde{H}_1\otimes \widetilde{H}_2 \otimes \widetilde{H}_3.$\\
Conversely, if $|\widetilde{\psi}\r$ can be transferred to $|\widetilde{\phi}\r$  under $\widetilde{H}_1\otimes \widetilde{H}_2 \otimes \widetilde{H}_3$, the LU equivalence of $|\psi\r$ and $|\phi\r$ can be easily seen.\qed
\bigskip
\noindent{\bf Remark:} Since $|\psi\r,|\phi\r$ are HOSVD states, we can choose $H_m=S^{(m)}$.  Then the unitary transformations from $M_{\psi, m}$ to $M_{\phi, m}$ must lie in $H_m^M=diag(S^{(m)}, ..., S^{(m)})$.
Clearly,  $\widetilde{H}_m\subseteq H_m$. From this point of view, our method can make the symmetry group into smaller blocks than the HOSVD decomposition technique.

\bigskip
\noindent{\bf Example \ 1.}
 Let
 {
 \small
 $|\psi\rangle=\sqrt{\frac{1}{6}}|111\rangle+\sqrt{\frac{1}{4}}|123\rangle+\sqrt{\frac{1}{12}}|132\rangle
 +\sqrt{\frac{1}{8}}|212\rangle+\sqrt{\frac{1}{24}}|221\rangle+\sqrt{\frac{1}{3}}|233\rangle$
 }
 be a pure state in $\mathbb{C}^2\otimes\mathbb{C}^3\otimes\mathbb{C}^3$\\

By the method of Bin Liu, we get that
{
\small
$${|\psi\rangle}_1{|\psi\rangle}_1^\dag= \left[
                                         \begin{array}{cc}
                                           \frac{1}{2} &   \\
                                            & \frac{1}{2}  \\
                                             \end{array}
                                             \right],$$
$${|\psi\rangle}_2{|\psi\rangle}_2^\dag= \left[
                                         \begin{array}{ccc}
                                           \frac {7}{24}&  &  \\
                                            & \frac {7}{24}&  \\
                                            & & \frac {5}{12} \\
                                             \end{array}
                                             \right],$$
$${|\psi\rangle}_3{|\psi\rangle}_3^\dag= \left[
                                         \begin{array}{ccc}
                                           \frac {5}{24}&  &  \\
                                            & \frac {5}{24}&  \\
                                            & & \frac {7}{12} \\
                                             \end{array}
                                             \right].$$

}

So  $|\psi\rangle$ itself is a HOSVD state. By the HOSVD decomposition, $S^{(1)}\otimes S^{(2)}\otimes S^{(3)}$ has the following form
{
\small
$$U(2)\otimes
\left[
  \begin{array}{cc}
    U(2) &  \\
     & e^{ i \theta_1} \\
  \end{array}
\right]\otimes
\left[
  \begin{array}{cc}
    U(2) &  \\
     & e^{i \theta_2 }\\
  \end{array}
\right].$$
}

The LU equivalent problem of $|\psi\rangle$  and the other state $|\phi\rangle$ reduce to whether there is a solution in $S^{(1)}\otimes S^{(2)}\otimes S^{(3)}$ transfers $|\psi\rangle$ to a core state of $|\phi\rangle$.

By our method, we calculate the three matrices $M_{\psi, 1}, M_{\psi, 2}, M_{\psi, 3} $ defined above instead of ${|\psi\rangle}_1{|\psi\rangle}_1^\dag, {|\psi\rangle}_2{|\psi\rangle}_2^\dag, {|\psi\rangle}_3{|\psi\rangle}_3^\dag$.
{
\small
$$M_{\psi, 1}=diag\left[\left[
     \begin{array}{cc}
       \frac{5}{12} &  \\
        & \frac{1}{6}\\
     \end{array}
   \right],
   \left[
     \begin{array}{cc}
       \frac{1}{12} &  \\
        & \frac{1}{3}\\
     \end{array}
   \right],
   \left[
     \begin{array}{cc}
       \frac{1}{4} &  \\
        & \frac{1}{6}\\
     \end{array}
   \right],
   \left[
     \begin{array}{cc}
       \frac{1}{4} &  \\
        & \frac{1}{3}\\
     \end{array}
   \right]\right],$$

$$M_{\psi, 2}=diag
\left[
\left[
  \begin{array}{ccc}
    \frac{7}{24} &  &  \\
     & \frac{7}{24} &  \\
     &  & \frac{5}{12} \\
  \end{array}
\right],
\left[
  \begin{array}{ccc}
    \frac{7}{24} &  &  \\
     & \frac{1}{24} &  \\
     &  & \frac{1}{12} \\
  \end{array}
\right],
\left[
  \begin{array}{ccc}
    0 &  &  \\
     & \frac{1}{4} &  \\
     &  & \frac{1}{3} \\
  \end{array}
\right]
\right],$$

$$M_{\psi, 3}=diag
\left[
\left[
  \begin{array}{ccc}
    \frac{5}{24} &  &  \\
     & \frac{5}{24} &  \\
     &  & \frac{7}{12} \\
  \end{array}
\right],
\left[
  \begin{array}{ccc}
    \frac{5}{24} &  &  \\
     & \frac{1}{8} &  \\
     &  & \frac{1}{4} \\
  \end{array}
\right],
\left[
  \begin{array}{ccc}
    0 &  &  \\
     & \frac{1}{12} &  \\
     &  & \frac{1}{3} \\
  \end{array}
\right]
\right].$$
}

Actually all the three matrices are  canonical forms of themselves under their direct groups $H_1^M, H_2^M, H_3^M$.\\
The direct group $\widetilde{H}_1\otimes \widetilde{H}_2 \otimes \widetilde{H}_3$ of the canonical form $|\widetilde\psi\r$  has the following form
{
\small
$$
\left[
  \begin{array}{cc}
    e^{ i\theta_1} &  \\
     & e^{ i\theta_2} \\
  \end{array}
\right]
\otimes
\left[
  \begin{array}{ccc}
    e^{i\theta_3} &  &  \\
     & e^{i\theta_4} &  \\
     &  & e^{i\theta_5} \\
  \end{array}
\right]
\otimes
\left[
  \begin{array}{ccc}
    e^{ i\theta_6 }&  &  \\
     & e^{ i\theta_7} &  \\
     &  & e^{ i\theta_8} \\
  \end{array}
\right].
$$
}
Hence the problem to decide whether $|\psi\r$ is LU equivalent to the other state $|\phi\r$
can be reduced to  whether there is a solution in $\widetilde{H}_1\otimes \widetilde{H}_2 \otimes \widetilde{H}_3$ transfers $|\widetilde{\psi}\r$ to $|\widetilde{\phi}\r$. It is an easy problem
(\cite{bliu}).

\bigskip
\noindent{\bf Example \ 2.}
 Let
 {\small
 $|\psi\rangle=\sqrt{\frac{2}{15}}|113\r+\sqrt{\frac{1}{6}}|121\r+\sqrt{\frac{1}{15}}|132\r
 +\sqrt{\frac{1}{5}}|212\r+\sqrt{\frac{1}{15}}|223\r+\sqrt{\frac{1}{10}}|231\r
 +\sqrt{\frac{1}{15}}|311\r+\sqrt{\frac{1}{15}}|323\r+\sqrt{\frac{2}{15}}|333\r$
 }
 be a pure state in $\mathbb{C}^3\otimes\mathbb{C}^3\otimes\mathbb{C}^3$.\\

Here we also use Bin Liu's method first.

After we calculating the following matrices
{
\small
 $${|\psi\rangle}_1{|\psi\rangle}_1^\dag=
\left[
  \begin{array}{ccc}
    \frac{11}{30} &  &    \\
     & \frac{11}{30} &    \\
     & & \frac{4}{15}   \\
  \end{array}
\right],
$$
$${|\psi\rangle}_2{|\psi\rangle}_2^\dag=
 \left[
  \begin{array}{ccc}
    \frac{2}{5} &  &    \\
     & \frac{3}{10} &    \\
     & & \frac{3}{10}   \\
  \end{array}
\right],
$$

$${|\psi\rangle}_3{|\psi\rangle}_3^\dag=
\left[
  \begin{array}{ccc}
    \frac{1}{3} &  &    \\
     & \frac{1}{3} &    \\
     & & \frac{1}{3}   \\
  \end{array}
\right],
$$
}
we can see that $|\psi\r$ itself is a HOSVD state. By the HOSVD decomposition, $S^{(1)}\otimes S^{(2)}\otimes S^{(3)}$ has the following form
$$
\left[
  \begin{array}{cc}
    U(2) &  \\
     & e^{ i \theta_1} \\
  \end{array}
\right]\otimes
\left[
  \begin{array}{cc}
    e^{i \theta_2 }&  \\
     & U(2) \\
  \end{array}
\right]\otimes
U(3).$$

That is, if we use Bin Liu's method, the LU equivalent problem of $|\psi\rangle$  and the other state $|\phi\rangle$ reduce to whether there is a solution in $S^{(1)}\otimes S^{(2)}\otimes S^{(3)}$ transfers $|\psi\rangle$ to the core state of $|\phi\rangle$. But this is also a complicated problem.

Now we use our method. We calculate the three matrices $M_{\psi, 1}, M_{\psi, 2}, M_{\psi, 3} $ defined above

{\small

$$M_{\psi, 1}=diag
\left[
\left[
  \begin{array}{ccc}
    \frac{11}{30} &  &    \\
     & \frac{11}{30} &    \\
     & & \frac{4}{15}   \\
  \end{array}
\right]
,
 \left[
  \begin{array}{ccc}
    \frac{2}{15} &  &    \\
     & \frac{1}{5} &    \\
     & & \frac{1}{15}   \\
  \end{array}
\right],
 \left[
  \begin{array}{ccc}
    \frac{7}{30} &  &    \\
     & \frac{1}{6} &    \\
     & & \frac{1}{5}   \\
  \end{array}
\right]\right],$$

$$M_{\psi, 2}=diag
\left[
 \left[
  \begin{array}{ccc}
    \frac{2}{5} &  &    \\
     & \frac{3}{10} &    \\
     & & \frac{3}{10}   \\
  \end{array}
\right]
,
 \left[
  \begin{array}{ccc}
    \frac{1}{15} &  &    \\
     & \frac{1}{15} &    \\
     & & \frac{2}{15}   \\
  \end{array}
\right],
 \left[
  \begin{array}{ccc}
    \frac{1}{3} &  &    \\
     & \frac{7}{30} &    \\
     & & \frac{1}{6}   \\
  \end{array}
\right]\right],$$

$$M_{\psi, 3}=diag
\left[
\left[
  \begin{array}{ccc}
    \frac{4}{15} &  &    \\
     & \frac{4}{15} &    \\
     & & \frac{1}{5}   \\
  \end{array}
\right],
 \left[
  \begin{array}{ccc}
    \frac{1}{15} &  &    \\
     & \frac{1}{15} &    \\
     & & \frac{2}{15}   \\
  \end{array}
\right],
\left[
  \begin{array}{ccc}
    \frac{1}{15} &  &    \\
     & \frac{1}{5} &    \\
     & & \frac{2}{15}   \\
  \end{array}
\right],
\left[
  \begin{array}{ccc}
    \frac{4}{15} &  &    \\
     & \frac{2}{15} &    \\
     & & \frac{1}{5}   \\
  \end{array}
\right]\right].$$
}
Clearly, the three matrices are  canonical forms of themselves under their direct groups $H_1^M, H_2^M, H_3^M$. The direct group $\widetilde{H}_1\otimes \widetilde{H}_2 \otimes \widetilde{H}_3$ of
the canonical form $|\widetilde{\psi}\r$ has the following form
{\small

$$
\left[
  \begin{array}{ccc}
    e^{i\theta_1} &  &  \\
     & e^{i\theta_2} &  \\
     &  & e^{i\theta_3} \\
  \end{array}
\right]
\otimes
\left[
  \begin{array}{ccc}
    e^{i\theta_4} &  &  \\
     & e^{i\theta_5} &  \\
     &  & e^{i\theta_6} \\
  \end{array}
\right]
\otimes
\left[
  \begin{array}{ccc}
    e^{ i\theta_7 }&  &  \\
     & e^{ i\theta_8} &  \\
     &  & e^{ i\theta_9} \\
  \end{array}
\right].
$$
}

Hence the problem to decide whether $|\psi\r$ is LU equivalent to another state $|\phi\r$
can be reduced to  whether there is a solution in $\widetilde{H}_1\otimes \widetilde{H}_2 \otimes \widetilde{H}_3$ transfers $|\widetilde{\psi}\r$ to $|\widetilde{\phi}\r$.\\

From the above two examples, it can be seen that our method is better than Bin Liu's.

\bigskip

\section{ Multipartite pure states }

Our method can be generalized to multipartite pure states. 
Given a $N$-partite pure HOSVD state
$$|\psi\rangle\in\mathbb{C}^{I_1}\otimes \mathbb{C}^{I_2}\otimes...\otimes \mathbb{C}^{I_N},$$
we can suppose that
$$|{\psi}\rangle=\sum\limits_{k=1}^{t^{(i)}}{\sum\limits_{l=1}^{\mu_k^{(i)}}{\lambda_{k}^{(i)}}}|{n_l^{(i), k}\rangle_i
\otimes|{v_{l}^{(i), k}}\rangle}_{\neg i},$$
where $ i=1, 2, ..., N,$ $n_l^{(i), k}\doteq l+\sum\limits_{s=1}^{k-1}{\mu_s^{(i)}}$ and $|{v_{l}^{(i),k}}\rangle_{\neg i}$ is just the normal vector when we collect the $n_l^{(i), k}$-th term of the i-th system of $|\psi\r$. For each $1\leq i\leq N$, $|\psi\rangle$ can be looked as the purify of the mixed $N-1$ partite state
$$\rho^{(i)}=\sum\limits_{k=1}^{t^{(i)}}{{\lambda_{k}^{(i)}}^{2}}{\sum\limits_{l=1}^{\mu_k^{(i)}}|{v_{l}^{(i), k}\rangle
_{\neg i\neg i}\langle{v_{l}^{(i), k}|}}}.
$$
Factoring out the k-th part of $\rho^{(i)}$ ,
$$\rho_k^{(i)}={{\lambda_{k}^{(i)}}^{2}}{\sum\limits_{l=1}^{\mu_k^{(i)}}|{v_{l}^{(i), k}\rangle_{\neg i\neg i}\langle{v_{l}^{(i), k}|}}}.
$$
Purifying $\rho_k^{(i)}$, we get a $N$-partite pure state
$$|\psi ^{(i), k}\rangle=\sum\limits_{l=1}^{\mu_k^{(i)}}|{n_l^{(i), k}\rangle_i\otimes|{v_{l}^{(i), k}}\rangle}_{\neg i}.$$
Considering the m-mode of $|\psi ^{(i), k}\rangle$, we have the following matrices
$$ M_{\psi, m}^{i, k}=|\psi ^{(i), k}\rangle_m|\psi ^{(i), k}\rangle_m^\dag, 1\leq m\neq i\leq N, k=1, ..., t^{(i)}.$$
Combined all this matrices into one large matrix, we get a matrix

 $$M_{\psi, m}=diag(M_{\psi,m}^{1, 1}, ..., M_{\psi, m}^{1,t^{(1)}},...,\widehat{M_{\psi, m}^{m, 1}, ..., M_{\psi, m}^{m, t^{(m)}}},
..., M_{\psi, m}^{N, 1}, ..., M_{\psi, m}^{N, t^{(N)}}),$$
where the cap means that the matrices under the cap should be deleted from the set of matrices. We can see that there
are $L_m=\mathbb{}\sum\limits_{i=1, i\neq m}^N t^{(i)}$ blocks of matrices in matrix $M_{\psi, m}$.

Let $H_m^M$ be  the direct group of $M_{L_m\times I_m}(\mathbb{C})$ which is defined by size $ \{r_k=I_m\}_{k=1}^{L_m}$ and restricted by $E_1=\{1,2,...,L_m\}$. Given a matrix $M_{\psi, m} $, by the algorithm in \cite{Heydar}, we have a matrix $U_{\psi, m}^{\widetilde{\psi}, M}$ transfers $M_{\psi, m} $ to its canonical form $M_{{\psi}, m}^0 $
with invariant subgroup $ \widetilde{H}_m^M=diag(\underbrace{\widetilde{H}_m, ..., \widetilde{H}_m}_{L_m})$.
Suppose
$$U_{\psi, m}^{\widetilde{\psi}, M}= diag(\underbrace{U_{\psi, m}^{\widetilde{\psi},  m}, ..., U_{\psi, m}^{\widetilde{\psi}, m}}_{L_m})\in H_m^M$$
with $U_{\psi, m}^{\widetilde{\psi}, m}$ in the $I_m$-th sequential principal minor.

A state
\begin{equation}
|\widetilde{\psi}\rangle \doteq U_{\psi, 1}^{\widetilde{\psi}, 1}\otimes U_{\psi, 2}^{\widetilde{\psi}, 2}\otimes...\otimes U_{\psi, N}^{\widetilde{\psi}, N}|\psi\rangle.
\end{equation}
is called a reduce form of $|\psi\rangle$, where  all the unitary matrices $\{U_{\psi, m}^{\widetilde{\psi}, m},1\leq m\leq N$\} can be calculated as above.

\bigskip

\noindent{\bf Theorem  \ 3.} Let $|\psi\rangle, |\phi\rangle$ be HOSVD states.
Then $|\psi\rangle$ and $|\phi\rangle$ are LU equivalent if and only if $|\widetilde{\psi}\rangle$
can be transferred to $|\widetilde{\phi}\rangle$ under $\widetilde{H}_1\otimes \widetilde{H}_2\otimes... \otimes \widetilde{H}_N$.

The proof is just similar with theorem 2.

\section {conclusion}

In this paper, we consider the problem of LU equivalence. After analysing the necessary conditions  of the LU equivalence, we obtain the conclusion of the simultaneously unitary equivalence of two order sets of matrices. A reduce form for each quantum state is obtained by virtue of  the algorithm for dealing with the unitary equivalence under direct group. Accordingly, the LU equivalent problem can be reduced into a simpler one. Our method is more efficient than that in \cite{bliu}. Hence the algorithm for unitary equivalence under direct group is an important way to study the LU equivalent problem.

\bigskip
$Acknowledgments:$ We would like to thank Xuena Zhu and Pengfei Guo for helpful discussions. We are very grateful to the referee for carefully reading and helpful comments.

\end {document}